# Response to "Reply to comment on 'Divergent and Ultrahigh Thermal Conductivity in Millimeter-Long Nanotubes'"

Qin-Yi Li[1,2]

[1]Department of Aeronautics and Astronautics, Kyushu University, Fukuoka 819-0395, Japan
[2]International Institute for Carbon Neutral Energy Research (WPI-I2CNER), Kyushu University, Japan

More than one year ago, Prof. Chih-Wei Chang and the co-authors published "Divergent and Ultrahigh Thermal Conductivity in Millimeter-Long Nanotubes" [1] in PRL and we submitted a comment [2]. After some while we received Prof. Chang et al.'s reply, which is almost the same as the arXiv preprint [3], and responded to the reply promptly.

There was no further reply from Prof. Chang in the following several months and the PRL editor consulted a third-party referee who gave a report as follows:

"In their Comment, Li et al point out the deficiencies of the analysis of the data of the paper by Lee et al, PRL 118, 135901 (2017). I find their arguments persuasive and their improved analysis valid. As they point out that the better analysis leads to a factor of 2 difference to the value of thermal conductivity, the Comment should definitely be published. The Reply to the Comment does not address the issues raised in a satisfactory matter, as very well described by Li et al in their response to the Reply. Thus I recommend the publication of the Li Comment, but the Reply is not adequate, and should acknowledge the deficiencies of the original model. In addition, I found a gross error in the original paper, in the supplementary material, where the analysis is described in detail. This also leads to a large error in the results. Namely, in the derivation of eq (s4), the total thermal resistance of the heater beam was defined as R_bi=2L/(kappa_bi A). This is wrong. The conductance of one half of the beam (from center to the bath) is of course G=kappa_bi A/L, if the total length of the beam is 2L. Thus the total conductance of the two parallel heat paths is Gtot= G1+G2 = 2 kappa_bi A/L. The total thermal resistance is then the inverse of the conductance, i.e. L/(2 kappa_bi A), which agrees with Li et al's result Rbi = lb/(4lambda_b Ab), as lb in their formula is the total length (=2L). The authors' result is four times two high, leading to a wrong equation (S7) and thus final equation used to calculate kappa_12. The bottom line is, the data needs to be reanalyzed with modeling that Li et al suggest."

On the request of some readers, I personally post here the detailed response to "Reply to comment on 'Divergent and Ultrahigh Thermal Conductivity in Millimeter-Long Nanotubes' ".

# Response Letter

Dear Editor, Prof. Chang and the co-authors of "Divergent and Ultrahigh Thermal Conductivity in Millimeter-Long Nanotubes",

As a quick response to the authors' reply, I would like to point out: (1) **the measured heat flux and the temperature rise in the sensors are also quite small,** so we must use the exact model to extract the apparent thermal resistance even if the non-parabolic deviation in the heater temperature is less than 0.1%. **Lee et al.'s ignorance of this non-parabolic effect directly led to their surprising and doubtful results.** (2) What can be measured using the multi-probe method is **indeed the apparent thermal resistance between two probes**, which is definitely smaller than the heat conduction resistance of CNT due to radiation (regardless of the thermal contact resistance), thus the thermal conductivity is **overestimated** owing to radiation. Our calculation is actually very clear if the readers rigorously solve the differential equations, although we were not able to include the details in the PRL Comment due to the length limitation. The effect of radiation can also be roughly evaluated using a fin model, which can be found in a Heat Transfer textbook. In addition, all the other famous groups have adopted the *sectionally* parabolic temperature profile in the heater or the detailed simulation in the data analysis, **rather than assuming the parabolic temperature rise in the heater**. In the following, I will respond to each paragraph of the authors' reply and show the detailed calculation and analysis.

**Paragraph 1:** The proceeding comment by Li *et al*. [1] has two points: (1) the temperature profile of the heater is not parabolic, and (2) that radiation heat loss from the single-wall carbon nanotube (SWCNT) induces an overestimation of the thermal conductivity of the sample. Here we show that Li *et al*. have confused two different measurement methods and misidentified our method to be similar to theirs. Therefore (1) introducing the non-parabolic correction by Li *et al*. has negligible (<0.1%) effects to our results; and (2) our measurements in fact *underestimate* the thermal conductivity of the SWCNTs, as emphasized in our paper [2].

**Response:**

As clearly shown in Fig. 1 in our Comment (also put here as the following Fig. R1), what we have analyzed is the four-probe scheme in the authors' paper. Fig. 1(c) is the thermal resistance circuit with an emphasis on the heater. The heater should have a sectionally parabolic temperature profile no matter how many probes are used in the experiment. Even a small deviation from the parabolic temperature profile should be carefully considered since the temperature rise in the sensor and the measured heat flux were also very small. **A deep reason is that the parabolic profile just indicates no heat flow from the heater to the test sample according to the Fourier's law.**

We have explained the surprisingly high and divergent thermal conductivities by **rigorously solving differential equations** and the details will be presented in the following responses.

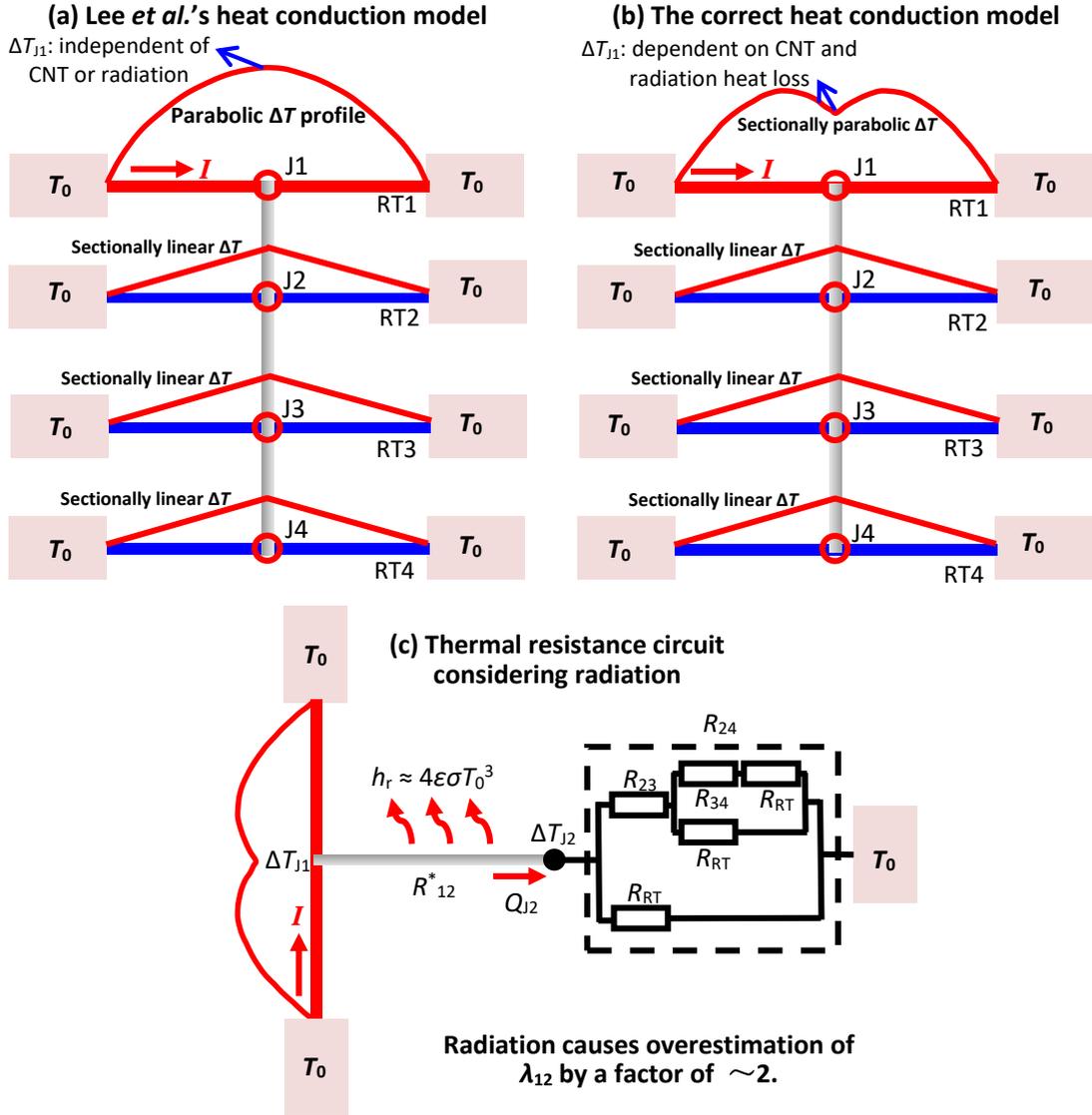

**Figure R1. Heat conduction models**

**Paragraph 2:** There are two different experimental methods for measuring thermal conductivity of a sample. First, one can supply a constant power ($P_h$) to a heater and then measure its temperature rise before ($\Delta T_{h,before}$) and after ($\Delta T_{h,after}$) connecting it to a sample. The measured thermal conductance ($K_m$) of the sample is obtained using

$$K_m = P_h \left( \frac{1}{\Delta T_{h,after}} - \frac{1}{\Delta T_{h,before}} \right) \quad (1)$$

As shown in Fig. 1(a), because only one probe is used for simultaneous heating and sensing, we dub it "one-probe method". Many scanning thermal microscopes [3,4], Prof. X. Zhang and Prof. K. Takahashi's previous works [5-8], optical techniques [9-11], and one of our previous works (see Methods I & II in Ref. [12]) have employed this method. Similar to Ref. [11], in which $\Delta T$ was measured using Raman shifts and $P_h$ was obtained from the laser absorption coefficient of a SWCNT, experiments using one-probe method commonly employ a "source $P_h$, measure $\Delta T$" measurement scheme.

**Response:**

Firstly, Eq. (1) is a wrong expression for the T-type method (or the "one-probe" method). In the T-type method, the measured average temperature rise of the heater/sensor is solved to be

$$\Delta T_1 = \frac{R_{b1}P}{12} + \frac{R_{b1}P}{4(R_{b1}/R_{app}+1)} \tag{R1}$$

where $P$ is the electrical power, $R_{b1}$ is the equivalent thermal resistance of the heater and equals $l_b/(4\lambda_b A_b)$ ($l_b$ = total probe length, $\lambda_b$ = probe thermal conductivity and $A_b$ = cross-sectional area of the probe) if the test sample is connected to the center of the heater, and $R_{app}$ is the apparent thermal resistance between the probe and the heat sink involving the heat conduction resistance of the test sample, thermal contact resistance and radiation effect. In fact, the apparent thermal resistance is directly extracted by linearly fitting the *T-P* curve by Eq. (R1) with the calibrated sensor properties rather than comparing the temperature rise before and after attaching the test sample. Anyway, if we rewritten the expression as a function of the temperature rise before and after attaching the sample with the same electrical heating power, the expression for apparent thermal conductance, $K_{app}$, should be given as

$$K_{app} = \frac{4P(\Delta T_{1,before} - \Delta T_1)}{3\Delta T_{1,before}(4\Delta T_1 - \Delta T_{1,before})} \tag{R2}$$

where the temperature rise under electrical power $P$ before sample attachment is expressed as

$$\Delta T_{1,before} = \frac{R_{b1}P}{3} \tag{R3}$$

(Note: here $R_{b1}$ is defined as $R_{b1} = l_b/(4\lambda_b A_b)$). Thus Eq. (1) in Lee et al.'s reply is simply wrong. Eq. (R3) was used by Lee *et al.* to analyze the temperature rise in the heater **after attaching the test sample**, which is wrong no matter how many probes are used in the experiment.

Secondly, I would like to point out that **the volumetric heat source due to Joule heating is different from the local (point) heat source in the heat conduction analysis**. The confusion of the point heat source and volumetric heat source can account for all the mistakes made in the authors' PRL paper and the reply. More relevant details will be shown in the next response.

**Paragraph 3:** On the other hand, many other experiments had incorporated an independent heater and an independent sensor, as displayed in Fig. 1(b), for measuring nanowires[13-15], nanotubes [16-18], or graphene [19,20]. Here $K_m$ is obtained using:

$$K_m = \frac{1}{\Delta T_h - \Delta T_s}\left(\frac{P_h \Delta T_s}{\Delta T_h + \Delta T_s}\right) = \frac{P_s}{\Delta T_h - \Delta T_s} \tag{2}$$

where $\Delta T_h$ and $\Delta T_s$ is temperature rise of the heater and the sensor, respectively. Note that the term in the bracket denotes the fraction of the total heater power received by the sensor. In our work, $\Delta T_h - \Delta T_s$ was kept constant and thermal current flowing through the sensor was measured (i.e. $P_s = 2K_s \Delta T_s$, where $K_s$ is the thermal conductance of the sensor beam). This method is dubbed "two-probe method", using a "source $\Delta T$, measure $P_s$" scheme. Unlike the one-probe method in which the heater must be located at the ends of a multiprobe device, the two-probe method has no such limitation.

**Response:**

**Lee et al.'s mistake is not that they used multiple probes, but that their data analysis is wrong which directly led to their surprising and doubtful results.** Here Eq. (2) in the Reply is also wrong, which should have been caused by **the confusion of point heat source and Joule volumetric heat source**.

Firstly, I would like to introduce a decent work by Prof. Li Shi's group (Smith, Brandon, et al. Advanced Materials (2016)), as shown in the following Fig. R2(a), where a four-probe method was used to measure the thermal conductivity of few-layer black phosphorous. **Please note** that (1) the temperature profile in the heater is sectional (the part of flat profile is caused by the line contact geometry) and (2) the electrical Joule heating is denoted as **volumetric $(IV)_1$** in the thermal resistance circuit, rather than the point heat source adopted in Lee et al.'s work (Fig. R2(b)).

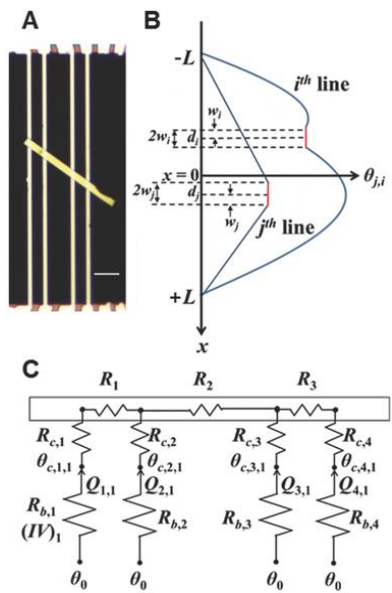

(a) Heat conduction model in Prof. Li Shi's work (Advanced Materials, 2016)

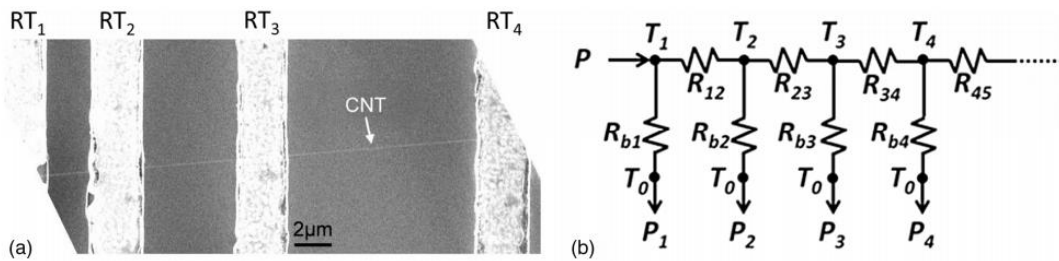

(b) Fig. (1) in Lee et al.'s paper, where a point heat source is used in the thermal circuit

Fig. R2 Heat conduction models for the four-probe method

Therefore, some famous groups have already used multiple probes to detect nanoscale heat conduction, but they used rigorous analysis to extract the thermal properties. Prof. Renkun Chen's group and Prof. Li Shi's group used exact solutions to differential equations to extract the apparent thermal resistance, while Prof. Eric Pop's group used detailed COMSOL simulation to extract thermal conductivity of graphene. Lee et al.'s mistake in the data analysis directly led to the surprisingly high and divergent thermal conductivity in CNT even for defected samples.

**Paragraph 4:** Now we discuss how would the radiation heat loss from a SWCNT make $K_m$ deviate from the intrinsic value of thermal conductance ($K$). Note that in Figs. 1(a & b), we always have $P_h > P_s$ whenever there is radiation heat loss from the sample. However, because the "source $P_h$, measure $\Delta T$" scheme is used in Fig. 1(a), it results in $K_m > K$. On the other hand, we had employed a "source $\Delta T$, measure $P_s$" scheme in Fig. 1(b), thus we concluded $K_m < K$ [2]. In fact, our method is equivalent to a two-probe electrical resistance measurement using a "source $V$, measure $I$" scheme, as shown in Fig. 1(c). Readers can easily verify our statements by analyzing the circuit.

**Response:**

Here I redraw the thermal resistance circuit in Fig. R3. Lee *et al.* extracted the thermal conductivity of a CNT segment using the following equation:

$$\lambda_{12} = \frac{Q_{J2} l_{12}}{A_{12} (\Delta T_{J1} - \Delta T_{J2})} \quad (R4)$$

where $\lambda_{12}$ denotes the thermal conductivity in the CNT segment between J1 and J2; $\Delta T_{J1}$ and $\Delta T_{J2}$ denote the temperature rise at the junctions between the CNT and RT1, RT2, respectively; $Q_{J2} = \Delta T_{J2} / R_{b2} + \Delta T_{J3} / R_{b3} + \Delta T_{J4} / R_{b4}$ is the measured heat flux at junction J2; $l_{12}$ and $A_{12}$ are the length and cross-sectional area of the CNT segment, respectively. **As discussed in our Comment, it should be noted that $\Delta T_{J1}$ is not a constant but will be lowered by radiation heat loss, which will cause overestimation of $\lambda_{12}$. In the end, what can be truly measured is the apparent thermal resistance between J1 and J2.**

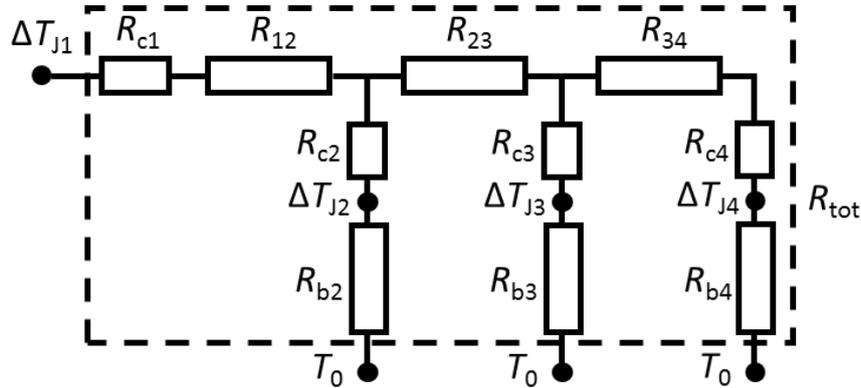

**Fig. R3 Thermal resistance circuit.** $R_{i(i+1)}$ denotes the apparent thermal resistance of the CNT segment between J$i$ and J($i$+1); $R_{ci}$ and $R_{bi}$ represent the thermal contact resistance at the junction $i$ and the thermal resistance of the thermometer $i$, respectively; $R_{tot}$ represents the equivalent thermal resistance of the circuit in the dash-line box.

In addition, the volumetric Joule heating is also conducted through the heater itself, so $P_h$ is indeed always larger than $P_s$. When there is radiation, the heat flux conducted from J1 (denoted as $Q_{J1}$) is actually larger than the no-radiation condition. $Q_{J1}$ is part of $Q_{J1}$ and the effect of radiation heat loss on $Q_{J2}$ in Eq. (R4) depends on the ratio of the heat conduction resistance to the radiation resistance. **So it is not always true that $Q_{J2}$ is underestimated.** Anyway, the effect of radiation on the measured thermal conductivity is quite clear because the measured apparent thermal resistance is smaller than the heat conduction resistance due to radiation.

**Paragraph 5:** Likewise, because the measured thermal conductance of our 1mm-long SWCNT is $1.77 \pm 0.15 \times 10^{-11}$ W/K and the thermal conductance of the heater beam is more than $2 \times 10^{-8}$ W/K, the non-parabolic correction to Eq. (2) is smaller than 0.1% and will not affect our conclusions.

**Response:**

Firstly, the calculation in Paragraph 5 is based on the point heat source, which is wrong as stated in the previous responses.

Next, I will show that even a small deviation from the parabolic temperature profile should be carefully considered since the temperature rise in RT2 and the heat flux through the test sample were also very small. Both the deviation from the parabolic profile and the temperature rise in RT2 were caused by the heat flow in the attached sample, so these two temperatures should be on the same order. Lee et al. did not give the temperature data of RT$_2$ in the PRL paper or Supplementary Materials, and the authors just claim that $\Delta T_1 \sim 20K \gg \Delta T_2 \gg \Delta T_3 \gg \Delta T_4$. For the simple case of not considering radiation, the heat flux through the CNT sample, i.e. $Q_{12}$, and the average temperature rise in the sensor RT2, i.e. $\Delta T_2$, can be calculated from the midpoint temperature gradient in the heater's sectionally parabolic temperature profile based on the Fourier's law. The expressions for $Q_{12}$ and $\Delta T_2$ are respectively given by Eqs. (R5) and (R6) (neglecting the thermal contact resistance and the temperature rise in RT3, RT4):

$$Q_{12} = \frac{P}{2} \frac{R_{b1}}{R_{b1} + R_{tot}} \quad (R5)$$

$$\Delta T_2 = \frac{P}{4} \frac{R_{b1} R_{b2}}{R_{b1} + R_{tot}} = \frac{P}{4} \frac{R_{b1}^2}{R_{b1} + R_{tot}} \quad (R6)$$

where $P$ is the Joule heating power, $R_{b1}$ (= $R_{b1}$) = $l_b/(4\lambda_b A_b)$ is the effective thermal resistance of the heater RT1, and $R_{tot}$ is the total thermal resistance between the junction J1 and the heat sink, as shown in Fig. R3. The non-parabolic temperature deviation in the heater is expressed as

$$\Delta T_1 - \Delta T_1^* = -\frac{P}{4} \frac{R_{b1}^2}{R_{b1} + R_{tot}} \quad (R7)$$

where $\Delta T_1^*$ denotes the average temperature rise of a parabolic profile. It turns out that the non-parabolic temperature reduction is exactly the same as the temperature rise in RT2. It is easy to understand since the leaked heat flux from the heater was finally conducted through the sensors. **Therefore, Lee et al.'s measured temperature rise in RT2 should be on the 0.01K order or less if the non-parabolic temperature deviation is less than 0.1% of 20 K.** This small $\Delta T_2$ indicates that the total thermal resistance between the junction J1 and the heat sink is much larger than $R_{b1}$, which is normal for a long SWCNT if the probe dimensions were not well designed. So in Lee et al.'s measurements, the temperature increase in the heater before and after attaching the test sample should have been very close, and the detected heat flux and temperature rise in RT2 should also have been negligible. Regardless of the sensitivity problem, the only way to accurately extract the apparent thermal resistance between J1 and J2 is to consider the small temperature change, which is possible since we have the high precision multimeters.

Finally, I will demonstrate in more detail that radiation heat loss accounts for overestimation of thermal conductivity in the ultralong CNT by ~2 times, **using the rigorous model considering volumetric Joule heating and the sectionally parabolic temperature rise in the heater**. Above all, what can be truly measured is **the apparent thermal resistance** between J1 and J2, which has been shown in the previous equations and has also been recognized by Prof. Li Shi at UT Austin. This apparent thermal resistance is lowered due to radiation, and thus the thermal conductivity of CNT is overestimated.

As described in our Comment, the effect of radiation can be roughly evaluated using a fin model, showing that the thermal conductivity could have been overestimated by ~2 times for a 1mm-long sample. We further rigorously solved differential equations to evaluate the effect of radiation. Herein, we consider radiation heat loss in the long CNT segment between junctions J1 and J2 and neglect radiation heat loss from the short CNT segments between J2 and J4 as well as all the contact thermal resistance, as illustrated in Fig. R1(c). I would like to skip the lengthy differential equations and boundary conditions. In short, we analytically solved heat conduction equations to obtain the expressions for $R_{tot}$, $\Delta T_{J1}$, $\Delta T_{J2}$ and $Q_{J2}$, and finally deduced the ratio of Lee *et al.*'s measured thermal conductivity to the true value, $\lambda_{12,\text{exp}}/\lambda_{12,\text{true}}$, as follows,

$$\frac{\lambda_{12,\text{exp}}}{\lambda_{12,\text{true}}} = \frac{Q_{J2} l_{12}}{\lambda_{12,\text{true}} A_{12} \left( \Delta T_{J1}^* - \Delta T_{J2} \right)}$$

$$= \frac{R_{24}}{l_{12}/\left(\lambda_{12,\text{true}} A_{12}\right)} \left\{ \left(1 + \frac{R_{b1}}{4 R_{tot}}\right) \left[ \frac{\sinh(m l_{12})}{R_{24} \sqrt{h_r C_{12} \lambda_{12,\text{true}} A_{12}}} + \cosh(m l_{12}) \right] - 1 \right\} \quad (R8)$$

where $\Delta T_{J1}^*$ is the heater's midpoint temperature rise adopted by Lee et al, $h_r \approx 4\varepsilon\sigma T_0^3$ is the radiation heat transfer coefficient, $C_{12}$ is the circumference of CNT's cross section, and $R_{24}$ is the overall thermal resistance of the circuit in the dash-line box in Fig. R1(c). We take the lengths of CNT segments between J2, J3 and J4 as 10μm, in which case the radiation heat loss between J2 and J4 is negligible; $l_{12} = 1$mm, $\lambda_{12} = 4000$ W/mK, $\lambda_{23} = \lambda_{34} = 3000$ W/mK and $\lambda_b = 50$ W/mK. The other relevant parameters are taken from Lee *et al.*'s paper. In this case, Lee *et al.*'s measured thermal conductivity in a 1mm-long SWCNT is calculated to be 1.92 times of the true value at room temperature.

**Paragraph 6:** Lastly, regarding to Liu *et al*.'s experiment that suggests a convergent thermal conductivity of SWCNTs at length ($L$) ~10μm [11], we must point out that the absence of error bars in their data has much perplexed us. Because the same technique demonstrated by other groups have shown to exhibit >10% uncertainty in the temperature measurement [10,21,22], and, additionally, complex position dependent variations [10,22], these would render Liu *et al*.'s data inconclusive for $L > 5$μm.
**Response:**

If the position dependence is significant due to defects, then Lee et al.'s finding that the ultrahigh thermal conductivity holds for defected samples would be quite doubtful.

In conclusion, Lee et al.'s finding is surprising and doubtful. It turns out that their heat conduction model was not accurate and that radiation heat loss account for the overestimation in thermal conductivity. There has already been some benchmark studies and good reviews on the thermal conductivity in CNT (see "Marconnet A M, Panzer M A, Goodson K E. Thermal conduction phenomena in carbon nanotubes and related nanostructured materials. Reviews of Modern Physics, 2013, 85(3): 1295"). The scattering at the defect and the ~3500 W/mK value for SWCNT have been widely recognized. Therefore, the > 10,000 W/mK divergent thermal conductivity in SWCNT must be taken with care.

Best regards,
Qinyi Li, Dr. Eng.
Kyushu University, Japan
Email: qinyi.li@aero.kyushu-u.ac.jp